%% file: exact_solution_JPA.tex
\documentclass[12pt]{iopart}
\usepackage{epsfig}
\usepackage{subfigure}
\usepackage{pstricks,pst-node}
\usepackage{cases}
\usepackage{url}

\newcommand{\text}[1]{\textrm{\ #1\ }}
\newcommand{\elabel}[1]{\label{#1}}
\newcommand{\flabel}[1]{\label{#1}}
\newcommand{\ave}[1]{{\left\langle #1 \right\rangle}}
\newcommand{\aves}[1]{{\left\langle\{ #1 \}\right\rangle}}

\newcommand{\GC}{\mathcal{G}}

\newcommand{\OP}{\mathbf{O}}
\newcommand{\mult}{\otimes}
\newcommand{\idmat}{\mathbf{1}}

\newcommand{\Dtilde}{\tilde{D}}
\newcommand{\gcomment}[1]{}

\newcommand{\erf}{\mathcal{E}}
\newcommand{\gen}[2]{Q_{#1,#2}}
\newcommand{\tgen}[2]{\tilde{Q}_{#1,#2}}
\newcommand{\src}[2]{S_{#1,#2}}
\newcommand{\bra}[1]{\left\langle#1\right|}
\newcommand{\braket}[2]{\left\langle#1|#2\right\rangle}
\newcommand{\ket}[1]{\left|#1\right\rangle}
\newcommand{\WP}{\widetilde{\psi}}
\newcommand{\OC}{{\mathcal{O}}}
\newcommand{\bpma}{\left(\begin{array}{cc}}
\newcommand{\epma}{\end{array}\right)}
\newcommand{\bpmv}{\left(\begin{array}{c}}
\newcommand{\epmv}{\end{array}\right)}
\newcommand{\binom}[2]{{#1 \choose #2}}

\begin{document}
\title{Exact solution of the totally asymmetric Oslo model}
\author{Gunnar Pruessner$\dagger$}
\address{$\dagger$Department of Mathematics,
Imperial College London,
180 Queen's Gate,
London SW7~2BZ,
UK}
\ead{gunnar.pruessner@physics.org} 
\date{20 February, 2004}

\begin{abstract}
Recently it has been found 
[G. Pruessner and H.~J. Jensen, Phys.~Rev.~Lett. {\bf 91},  244303  (2003)]
that a totally
asymmetric variant of the Oslo model 
[K. Christensen {\it et~al.}, Phys.~Rev.~Lett. {\bf 77},  107  (1996)]
represents the entire universality class of the Oslo model with
anisotropy. The totally asymmetric model can be solved without scaling
assumptions by finding recursively the eigenvectors of the Markov matrix,
which can be suitably modified to produce the moment generating function
of the relevant observable. This method should be applicable to many
other stochastic processes.
\end{abstract}

\submitto{\JPA}


\maketitle


\section{Introduction}
Self-organised criticality (SOC) was originally introduced
\cite{BakTangWiesenfeld:1987} as an approach to understand $1/f$-noise
as well as the apparent abundance of power laws in nature, which is
generally accepted as the sign of scale-invariance. The idea is that
under very general circumstances driven stochastic processes develop
into a scale-invariant state without the
explicit tuning of parameters, contrary to what one would expect from
equilibrium critical phenomena \cite{Stanley:71}.

A very large zoo of SOC models has been developed \cite{Jensen:98}, with each
model having certain special features. However, based on large scale
numerical simulations it has become increasingly clear that many of the
models formerly thought of as representatives of entire universality
classes or even paradigms for a specific type of model, are either
not scale-invariant or at least do not follow simple scaling
\cite{BoulterMiller:2003,Grassberger:2002,JensenPruessner:2002b,DattaChristensenJensen:2000}.
In fact, models of SOC are notorious for slow convergence and deviations
from the expected behaviour.

Some models, however, show all features one would expect from a
``self-organised critical'' system: Consistent exponents and scaling,
universality, crossover between different classes etc. One of these
models is the so-called Oslo model \cite{ChristensenETAL:1996}, which
was motivated by an experiment \cite{FretteETAL:1996}. In a recent
paper \cite{JensenPruessner:2003}, it has been shown that any (small) amount of anisotropy will
drive this model eventually (in the thermodynamic limit) towards another
``fixed point'', which is represented by the ``totally asymmetric Oslo
model'' (TAOM). This model is solvable directly on the lattice without
making any scaling assumptions. Consequently, it is not only possible
to derive exponents, but also to calculate amplitudes of the moments of the
relevant observable.

The TAOM is totally asymmetric in the sense that particles can move in
one direction only, similar to the totally asymmetric exclusion process
(TASEP)
\cite{DerridaDomanyMukamel:1992,DerridaEvansHakimPasquier:1993}. The
TASEP has been solved using a matrix product state ansatz
\cite{DerridaEvansHakimPasquier:1993,DerridaEvans:1997}, so that it
seems reasonable to apply similar techniques to the present model. However,
there is a crucial difference between these two stochastic processes:
The relevant observables in the TASEP exist on a microscopic timescale,
i.e. there is an intrinsic timescale in the time-evolution of the
microstate of the system. In contrast, in the TAOM the relevant
observables are obtained by any dynamics which comply with a certain set
of rules. In that sense, the specific (microscopic) dynamics of the TAOM are
irrelevant. This is reflected in its theoretical treatment, in that the
TASEP is updated homogeneously (all sites evolve equally) but the TAOM
is perturbed once and is only observed after it is fully relaxed
(separation of time-scales).

In the following, the model is defined in terms of rules on a
lattice. Using a Markov matrix approach it is then solved and exponents
and amplitudes derived. After mapping it to a reaction-diffusion process
as well as various other processes, a more accessible continuum theory
is described.

\section{The Model}\label{sec:model}
The model is defined on a one-dimensional lattice of size $L$, where
each site $i=1,2,\dots,L$ has assigned a slope $z_i\in\{1,2\}$ and a critical slope
$z^c_i\in\{1,2\}$. From a flat initial configuration $z_i \equiv 1$ and
$z^c_i$ random, where $z^c_i=1$ is chosen with probability $p$ and $z^c_i=2$ otherwise,
the model evolves according to the following rules: 
\begin{enumerate}
\item (Driving) Increase $z_1$ by one unit (``initial kick'').
\item (Toppling) If there is an $i$ where $z_i > z^c_i$, decrease $z_i$
 by $1$ unit, $z_i\to z_i-1$, and increase the right nearest neighbour
 $j=i+1$ by $1$, $z_j \to z_j + 1$ (charging). A new $z^c_i$ is chosen
 at random from $\{1,2\}$, where $z^c_i=1$ is chosen with probability $p$ and
 $z^c_i=2$ otherwise.
\item Repeat the second step until $z_i \le z^c_i$ (``stable'')
      everywhere. Then
 proceed with the first step. 
\end{enumerate}
During toppling, the right neighbour is charged of course only if it
actually exists, i.e. $j\le L$, otherwise the toppling site $i$ relaxes
without charging another site, so that a unit leaves the system. Apart
from this boundary condition, the TAOM differs from the original Oslo
model \cite{ChristensenETAL:1996} in redistributing only a unit to the
right, rather than one to each side.

It is important to note that the value of $z^c_i$ is determined only
\emph{after} a site has discharged. Thus, if a \emph{stable} site $i$ is
in state $z_i=1$, its value of $z^c_i$ could be randomly chosen in the
moment when it is needed, i.e. when the site is charged again. If a
\emph{stable} site $i$ is in state $z_i=2$, then $z^c_i$ has necessarily
the value $2$. When this site is charged, it will relax to $z_i=2$ again
and a new random $z^c_i$ is drawn. If that is $z^c_i=1$, then the site topples
again and ends up in state $z_i=1$, otherwise it remains in state
$z_i=2$.

If all sites are stable, i.e. $z_i\le z_i^c$ for all $i\in [1,L]$, a
configuration is fully described by the values of the $z_i$ alone; if
$z_i=2$ then $z_i^c=2$, otherwise $z^c_i$ is random and has not yet been
used in the dynamics.

The number of times the second rule is applied, that is the number of
topplings, is the avalanche size $s$. The fundamental observable one is
interested in is the probability density function of these sizes, $P(s)$,
which is expected to obey simple scaling above a fixed lower cutoff $s_l$,
\begin{equation} \elabel{def_expos}
 P(s) = a s^{-\tau} \GC(s/s_0) \ ,
\end{equation}
with $s_0 = b L^D$, $\GC$ the universal finite-size scaling function, and metric factors $a$ and $b$
\cite{PrivmanFisher:1984}. \Eref{def_expos} is
the \emph{definition} of the two exponents $\tau$ and $D$. It entails
that the moments $\ave{s^n}$ of $P(s)$ behave like
\cite{JensenPruessner:2003}
\begin{equation} \elabel{moment_scaling}
 \ave{s^n} = a (b L^D)^{1+n-\tau} g_n \ \text{for } 1+n-\tau>0
\end{equation}
with universal amplitudes $g_n$ \cite{JensenPruessner:2003}. Thus,
assuming \eref{def_expos} one can derive $\tau$ and $D$ from the
behaviour of any two moments. Below, the exponents $\gamma_n$ from
$\ave{s^n} \propto L^{\gamma_n}$ will be used; \Eref{moment_scaling}
therefore means
\begin{equation} \elabel{gamma_n}
\gamma_n=D(1+n-\tau) \ .
\end{equation}

The time series of avalanches, $s(t)$, itself is not Markovian, while
the sequence of stable configurations of the system, given by the vector
$(z_1, z_2, \cdots, z_L)$, is. Since two consecutive stable
configurations are not necessarily linked by a unique sequence of
topplings, the sequence of avalanche sizes is not uniquely determined by
the sequence of configurations. Nevertheless, in the form of a generating
function this ambiguity can be built into the Markov matrix operating on
the distribution vector of configurations, so that the avalanche size
distribution can be determined by means of this specially prepared
Markov matrix.

\subsection{Abelian property} \label{sec:abelian}
Put simply, if a model is Abelian \cite{Dhar:1999}, it means that the
order of updates is irrelevant for its statistical properties. It is
exceptionally simple to see this property here: Firstly, for the final
state of an individual
site there is no difference between a certain number of charges arriving
at once or arriving sequentially. Secondly, if a site topples, it
pours particles on its right neighbour, but it will never receive
anything back from the neighbour. So, if a site at $z=1$ has received
$3$ units, it topples at least twice, but for this site it does not make
any difference whether it first moves one unit over to the right
neighbour and waits until all sites to its right have relaxed, or
whether it moves all units at once, $2$ with probability $q\equiv 1-p$
(namely the probability to have $z_i^c=2$ after the second toppling) and
$3$ with probability $p$.

In this informal sense, the Abelian property allows the updating to run
from left to right, completely relaxing each site during a sweep. If
there is no toppling on site $i$, the avalanche has stopped and sites
$j>i$ do not need to be checked for the toppling condition $z_j>z_j^c$
at all. This procedure makes very efficient Monte Carlo simulations
possible. Moreover, it defines an activity $a_i$, which is the
total number of charges received at site $i$ during an avalanche. The
activity will be used in Sec.~\ref{sec:reaction_diffusion}.

\section{Markov matrix approach} \label{sec:markov}
The tensor product $\mult$ used here is explained in detail in
\cite{Hinrichsen:2000}. In particular it has the property
(provided that $A$, $B$, $A'$ and $B'$ have approppriate ranks)
\begin{equation} \elabel{tensor_product}
 (A \mult B) \odot (A' \mult B') = (A \odot A') \mult (B \odot B')
\end{equation}
where $\odot$ stands for the appropriate operator: it is a matrix
multiplication if $A$, $B$, $A'$ and $B'$ are matrices, it is a
multiplication of a matrix and a vector if $A$ and $B$ are matrices
and $A'$ and $B'$ are vectors or vice versa, or it is an inner product if
they are all vectors. In particular, in the latter case it is 
\begin{equation} \elabel{tensor_product_inner}
 (a \mult b) (a' \mult b') = (a a')(b b') \ .
\end{equation}

First, we consider a single site system, which can be in exactly two
states, so that its distribution of states can be represented by a
two-row vector. By convention, the upper row corresponds to $z=1$ and
the lower row to $z=2$. Three matrices are introduced, corresponding
to the three possible outcomes of a single initial kick.

The matrix $S$ corresponds to a unit being absorbed, i.e. the site is
in state $z_1=1$ and $z^c_1=2$, which occurs with probability $q$. After
the charge, the system is in state $z_i=2$. Similarly, $T$ corresponds
to a single toppling due to the charge and $U$ corresponds to a double
toppling:
\begin{equation}
 S=\bpma 0 & 0 \\ q & 0 \epma
\qquad
 T=\bpma p & 0 \\ 0 & q \epma
\qquad
 U=\bpma 0 & p \\ 0 & 0 \epma
\end{equation}
In the following, the aim is to find an expression for the moment
generating function of the avalanche size distribution. To this end,
each matrix is multiplied by an appropriate power of $x$, so that
evaluating at $x=1$ gives the usual Markov matrix of this process, and
deriving by $x$ before evaluating at $x=1$ multiplies each process by
the number of topplings occurring in it, and similarly for higher order
moments \cite{vanKampen:1992}.

It will be motivated only \emph{a posteriori} that a dissipation
process is required, say with probability $0\le\epsilon\le 1$; this
process corresponds to charging without changing the state, i.e. an
identity operation $\idmat$, the latter being the $2\times 2$ identity
matrix. The resulting single site operator is therefore
\begin{equation}
 \OP_1(x) = \epsilon \idmat + \delta\left(S + x T + x^2 U\right) =
\bpma
\epsilon + x \delta p & x^2 \delta p \\
\delta q              & \epsilon + x \delta q
\epma
\end{equation}
with $\delta\equiv 1-\epsilon$. The eigenvectors and eigenvalues of this matrix are found to be
\elabel{evecs_L1}
\begin{equation}
\begin{array}{rlrlrl}
\bra{e_\lambda(x)} &= \left(\frac{1}{x}, 1\right) \quad &
\ket{e_\lambda(x)} &= \bpmv xp \\ q \epmv \quad &
\lambda &= \epsilon + x \delta \\
\bra{e_\mu(x)} &= \left(-\frac{q}{x}, p\right) \quad &
\ket{e_\mu(x)} &= \bpmv -x \\ 1 \epmv  \quad &
\mu &= \epsilon 
\end{array}
\end{equation}
where $\OP_1$ acts on bra-vectors $\bra{}$ from the right and on
ket-vectors $\ket{}$ from the left. The vectors are normalised such
that
\begin{equation} \elabel{kronecker_1}
 \braket{e_a}{e_b} = \delta_{a,b}
\end{equation}
with $\delta_{a,b}$ denoting the Kronecker delta-function. In order to
distinguish vectors of different size, in the following they are often
marked with an index $L$ to indicate a size $2^L$.

$\OP_L(x)$ is the operator which adds a unit on site $i=1$ and relaxes
the entire lattice of size $L$. It is a matrix of size $2^L \times 2^L$
and defined as
\begin{equation}\elabel{def_OP}
 \OP_L(x) = \epsilon \idmat^{\mult L} +  
\delta \left( S \mult \idmat^{\mult (L-1)} + x T \mult \OP_{L-1}(x) + x^2 U \mult \OP_{L-1}^2(x) \right)
\end{equation}
again with a dissipation rate $\epsilon$, leaving the state
unchanged. The bracket multiplied by $\delta$ consists of three terms:
The first term charges the site without toppling and leaves the rest
of the system unchanged by operating with the identity $\idmat^{\mult
(L-1)}$. The second term corresponds to a single toppling, which
charges the remaining system of size $L-1$ once. This term is derived using the
identity
\begin{equation}\elabel{recurrence}
 \left(T \mult \idmat^{\mult (L-1)} \right) \left(\idmat \mult \OP_{L-1}(x)\right) =
 T \mult \OP_{L-1}(x) \ .
\end{equation}
The third term is a double toppling of the site, giving rise to a double
charge of the remaining system.

The Abelian property mentioned above (Sec.~\ref{sec:abelian}) can be
expressed as the commutator for two charges on a system of size $L$, one
at site $i=1$, the other one at site $1+L-L'$ with $L'$ being the
size of the subsystem starting from the site receiving the second charge, 
\begin{equation} \elabel{abelian}
 \OP_L(x) \left(\idmat^{\mult (L-L')} \mult \OP_{L'}(x)\right) = 
 \left(\idmat^{\mult (L-L')} \mult \OP_{L'}(x)\right) \OP_L(x)
\end{equation}
where of course $L\ge L'$. The tensor multiplication used on $\OP_{L'}$
and also in \eref{recurrence} ensures that both matrices have the same
rank; they are ``filled with identity'' where they do not
act. \Eref{abelian} simply states that it does not matter for the
statistics whether the leftmost site of a right subsystem of size $L'$
in a system of size $L$ is charged first, followed by the leftmost site
of the entire system, or vice versa. Due to the asymmetry in the
dynamics, it is clear that a system of size $L$, initially charged at
site $i$, has the same statistics as a system of size $L-i+1$, charged at
its leftmost site. It might be interesting, however, to formally prove
\Eref{abelian}, which should be easily feasible using established
methods \cite{MeesterRedigZnamenski:2001,Dhar:1999}.

The distribution of states at time $t$ is the vector $\ket{P_t}_L$,
which has rank $2^L$, each row corresponding to the probability for the
system to be in the state encoded by that row. The encoding follows from
the row ordering convention introduced above and the use of the tensor
product in \eref{def_OP}. 

For $x=1$ the operator $\OP_L(x)$ is simply the Markov matrix acting on
$\ket{P_t}_L$, producing the
distribution of states at time $t+1$ \cite{vanKampen:1992}
\begin{equation}
 \ket{P_{t+1}}_L = \OP_L(1) \ket{P_t}_L \ .
\end{equation}
There exists at least one eigenvector with eigenvalue $1$, which is
therefore a stationary distribution. If the eigenvectors represent a
complete basis and the modulus of all other
eigenvalues is less than unity, this stationary distribution is unique
and reached by \emph{any} initial distribution. The stationary
distribution, denoted $\ket{0}_L$, is the focus of the following
calculations. It is shown below that it is unique.

One very important bra-eigenvector with eigenvalue $1$ of $\OP_L(x=1)$
is  
\begin{equation} \elabel{evec_norm}
 \bra{0}_L \equiv (\underbrace{1,1,\dots,1}_{2^L \text{\scriptsize times}})
\end{equation}
by normalisation. As has been indicated above, for general $x$, the
operator $\OP_L(x)$ becomes a moment generating function of the
avalanche size, 
if sandwiched between $\bra{0}_L$ and the stationary distribution:
\begin{equation} \elabel{def_q}
 \gen{L}{n}(x;\epsilon) \equiv \bra{0}_L \OP^n_L(x) \ket{0}_L
\end{equation}
This can be seen from \eref{def_OP} containing an $x$ for every
toppling. When the operator acts on a distribution, for each transition from
one state to another a power of $x$ corresponding to the number of
topplings enters and is multiplied by the probability to be in the
initial state (given by the initial distribution) and the transition probability
given by the transition matrix. The function $\gen{L}{n}(x;\epsilon)$ for general
$n$ is then the generating function for avalanches caused by
$n=1,2,\dots$ initial kicks. In particular
\begin{equation} \elabel{sm}
 \ave{s^m}_L = \left. \left(x \frac{d}{dx} \right)^m\right|_{x=1} \gen{L}{1}(x;\epsilon) \ .
\end{equation}

The aim of the following calculations is to find the generating function
$\gen{L}{1}(x;\epsilon)$ or at least the moments generated by it.

\subsection{General eigenvectors and eigenvalues of $\OP_L(x)$}
It would be very helpful if $\OP_L(x)$ could be written in the form
\begin{equation} \elabel{operator_in_evecs}
 \OP_L(x) = \sum_{i=0}^{2^L-1} \ket{i(x)}_L \lambda_{L,i}  \bra{i(x)}_L \ ,
\end{equation}
where $\bra{i(x)}_L$ denote the left hand and $\ket{i(x)}_L$ the right
hand eigenvectors of $\OP_L(x)$ with eigenvalues $\lambda_{i,L}(x)$ and
$i=0\dots 2^{L}-1$. \emph{A priori} it is not clear whether these
vectors actually exist. In the following they are constructed and it is
shown that setting $\epsilon=0$ leads to fundamental
problems. 

Assuming that $\ket{i}_{L-1}$ is an eigenvector with  eigenvalue
$\lambda_{L-1,i}$ of $\OP_{L-1}(x)$, the definition of $\OP_L(x)$,
\Eref{def_OP}, gives for an arbitrary vector $\ket{e}_1$
\begin{equation}\fl
 \OP_L(x) \left(\ket{e}_1 \mult \ket{i}_{L-1}\right) = 
 \left[ 
 \left\{
\epsilon \idmat + \delta \left(S + x T \lambda_{L-1,i} + x^2 U
 \lambda_{L-1,i}^2 \right)\right\} 
 \ket{e}_1\right] \mult \ket{i}_{L-1}
\end{equation}
where $\ket{e}_1$ contains two elements such that $\ket{e}_1 \mult
\ket{i}_{L-1}$ is a vector of $2^L$ elements. The matrix in the curly
brackets is simply $\OP_1(x \lambda_{L-1,i})$. So, if $\ket{e}_1$ is 
either $\ket{e_\lambda(x \lambda_{L-1,i})}$ or $\ket{e_\mu(x
\lambda_{L-1,i})}$ from \eref{evecs_L1}, then $\ket{e}_1 \mult
\ket{i}_{L-1}$ is an eigenvector of $\OP_L(x)$ with eigenvalues
$\epsilon + \delta (x \lambda_{L-1,i})$ or $\epsilon$. Thus, based on
\eref{evecs_L1}, one can write the eigenvectors of $\OP_L(x)$ as
\begin{equation}
\elabel{evec_tree}
\begin{array}{rl} 
 \ket{i}_{L} &= \ket{e_\lambda(x \lambda_{L-1,i})} \mult \ket{i}_{L-1} \\
 \bra{i}_{L} &= \bra{e_\lambda(x \lambda_{L-1,i})} \mult \bra{i}_{L-1} \\
 \ket{i+2^{L-1}}_{L} &= \ket{e_\mu(x \lambda_{L-1,i})} \mult \ket{i}_{L-1} \\
 \bra{i+2^{L-1}}_{L} &= \bra{e_\mu(x \lambda_{L-1,i})} \mult \bra{i}_{L-1} 
\end{array}
\end{equation}
and the eigenvalues as 
\begin{equation}
\elabel{eval_tree}
\begin{array}{rl}
 \lambda_{L,i}&=\epsilon + x \delta \lambda_{L-1,i} \\
 \lambda_{L,i+2^{L-1}}&=\epsilon 
\end{array}
\end{equation}
both with $i=0,1,\dots,2^{L-1}-1$. To start the hierarchy, one defines
\begin{equation}
\begin{array}{rl}
 \ket{0}_{1} &= \ket{e_\lambda(x)} \\
 \bra{0}_{1} &= \bra{e_\lambda(x)} \\
 \ket{1}_{1} &= \ket{e_\mu(x)} \\
 \bra{1}_{1} &= \bra{e_\mu(x)} 
\end{array}
\end{equation}
and the eigenvalues as 
\begin{equation}
\elabel{init_eval_tree}
\begin{array}{rl}
 \lambda_{1,0}&=\epsilon + x \delta \\
 \lambda_{1,1}&=\epsilon 
\end{array}
\end{equation}

Now it is clear why the quantity $\epsilon$ was necessary: For
$\epsilon=0$ all but one eigenvalues vanish, which can be seen from the
hierarchy of eigenvalues obtained by iterating
\eref{eval_tree}. Therefore, if the vanishing eigenvalue of $L-1$ is plugged into
$\bra{e_\lambda}$ or $\bra{e_\mu}$ according to \eref{evec_tree}, the
result is undefined, as can be seen from \eref{evecs_L1}, so that the
bra-eigenvectors cease to exist.

The fact that all but one eigenvalues vanish for $\epsilon=0$ is very
deceptive. Assuming that any initial condition $\ket{P}$ can be written
in terms of the eigenvectors of $\OP_L(1)$, say $\sum a_i \ket{i}$, this
suggests $\OP_L(1) \ket{P} = \ket{0}$. This, however, is wrong,
because for vanishing $\epsilon$ the operator $\OP_L(x)$ cannot be
written in the form \eref{operator_in_evecs} for $L>1$. And it must be wrong,
because, for example, kicking an empty system once cannot make it produce the
stationary distribution.

If the eigenvectors of $\OP_{L-1}$ are linearly independent, then one
can show, using the construction \eref{evec_tree}, that the eigenvectors
of $\OP_{L}$ are linearly independent as well, provided that
$\ket{e_\lambda(x \lambda_{L-1,i})}$ and $\ket{e_\mu(x
\lambda_{L-1,i})}$ are linearly independent. This is not the case for
$\epsilon=0$ (see the ket vectors in \eref{evecs_L1} with $x=0$) and
this is the basic reason why $\epsilon\ne0$ is needed for the time
being. However, for any $\epsilon\ne 0$ one can apparently construct a
diagonalising matrix for $\OP_L$. Thus, it can be written in the form
\eref{operator_in_evecs}. Especially, the eigenvectors have the property
(by induction)
\begin{equation} \elabel{kronecker_L}
 \braket{i(x)}{j(x)}_L = \delta_{i,j}
\end{equation}
and as all $2^L$ eigenvectors are linearly independent, they must span
the whole space so that
\begin{equation} \elabel{sum_i_identity}
 \sum_{i=0}^{2^L-1} \ket{i}_L\bra{i}_L = \idmat \ .
\end{equation}
In the form \eref{operator_in_evecs} the operator can now be applied to a
stationary distribution to give
\begin{equation} \elabel{qn_from_evecs}
 \gen{L}{n}(x;\epsilon) = \sum_{i=0}^{2^L-1} \braket{0}{i(x)}_L  \lambda_{L,i}^n \braket{i(x)}{0}_L
\end{equation}

\subsubsection{The stationary distribution}
From \eref{evec_tree} the stationary distribution can be derived
immediately. It is the eigenvector with eigenvalue $1$ of
$\OP_L(1)$. Setting $x=1$ in \eref{eval_tree} it is clear that
$\lambda_{L,i}=1$ requires a $\lambda_{L-1,j}=1$, which, together with
\eref{init_eval_tree}, gives the unique $\lambda_{L,0}=1$ provided that
$\epsilon<1$. If $\epsilon=1$, then \emph{all} eigenvalues are $1$, but
still all eigenvectors are linearly independent and therefore span the
entire space, so that \emph{all} initial distributions are
stationary. This is not surprising because $\epsilon=1$ simply means
that any added particle immediately dissipates from the system, so that
adding a particle is in fact just the identity operation.

For $0\le \epsilon < 1$ the stationary distribution is unique and all
other eigenvalues have modulus less than $1$. The eigenvector
corresponding to eigenvalue $\lambda_{L,0}=1$ is, according to \eref{evec_tree},
\begin{subequations}
\begin{eqnarray}
 \bra{0}_L & =& \bra{e_\lambda(1)}^{\mult L} = (1,1)^{\mult L} \\
 \ket{0}_L & =& \ket{e_\lambda(1)}^{\mult L} = \bpmv p \\ q\epmv ^{\mult L} \elabel{stat_dist}
\end{eqnarray}
\end{subequations}
which is consistent with the notation for the stationary distribution
and the normalisation eigenvector introduced in \eref{def_q} and
\eref{evec_norm}. The last line, \Eref{stat_dist}, indicates that the
stationary state is a product measure, i.e. a state at one site does not
depend on the state on any other site. In fact the spatial correlation
function of sites $\{i_1,i_2,\dots\}$ can easily be calculated by
``dressing'' the states of the sites by appropriate powers of a variable
$x_i$, in order to obtain the generating function of
the correlators. The function
\begin{equation} \elabel{full_corr}
\fl
 C(x_1,x_2,\dots,x_L) = \bra{0}_L 
\bpmv p x_1 \\ q x_1^{-1} \epmv 
\bpmv p x_2 \\ q x_2^{-1} \epmv 
\dots
\bpmv p x_L \\ q x_L^{-1} \epmv 
= \prod_i^L (px_i + q x_i^{-1})
\end{equation}
is the generating function of the state-correlators, where state $1$
stands for $z=1$ and state $-1$ for $z=2$. The states have the useful
property that the joint contribution of two sites is $1$ if both sites
are in the same state and $-1$ otherwise. The average state is obtained
from
\begin{equation}
\fl
 x_i \left. \frac{d}{dx_i}\right|_{x_1,\dots,x_L=1} \ln\left(C(x_1,x_2,\dots,x_L)\right) =
     \left. \frac{d}{dx_i}\right|_{x_1,\dots,x_L=1} C(x_1,x_2,\dots,x_L) =
 p-q
\end{equation}
Correspondingly, the connected two point correlation function of sites
$i$ and $j$ is given by
\begin{subnumcases}{
x_j        \frac{d}{dx_j}
x_i \left. \frac{d}{dx_i}\right|_{x_1,\dots,x_L=1} 
\ln\left(C(x_1,x_2,\dots,x_L) \right)
=
}
4 p q & for $i=j$ \\
0 & otherwise
\end{subnumcases}
This confirms the absence of correlations and is fully consistent with
the expected variance of the state.

\subsection{The hierarchy of generating functions}
Using \eref{qn_from_evecs}, one can now calculate the generating
function $\gen{L}{n}(x;\epsilon)$, 
by plugging the hierarchy of eigenvectors \eref{evec_tree} and
eigenvalues \eref{eval_tree} into \eref{qn_from_evecs} and using the
properties of $\mult$, see \Eref{tensor_product}. For $n=1$ it is
\begin{eqnarray*}
 \gen{L}{1}(x;\epsilon) & = & \sum_{i=0}^{2^{L-1}-1} x \delta \lambda_{L-1,i} 
                  \braket{0}{i}_{L-1} \braket{0}{e_\lambda(\lambda_{L-1,i})}_1  
\braket{i}{0}_{L-1} \braket{e_\lambda(\lambda_{L-1,i})}{0}_1  \nonumber \\
&&+ \sum_{i=0}^{2^L-1} \epsilon 
                  \braket{0}{i(x)}_L \braket{i(x)}{0}_L \nonumber 
\end{eqnarray*}
where the term proportional to $\epsilon$ comes from the $\epsilon$ in
every $\lambda_{L,i}$ (see \Eref{eval_tree}). From \eref{sum_i_identity}
it is clear that the last sum gives $1$. The two projections give
\begin{equation*}
 \braket{0}{e_\lambda(\lambda_{L-1,i})}_1 = \lambda_{L-1,i} p + q 
\quad \text{and} \quad
\braket{e_\lambda(\lambda_{L-1,i})}{0}_1 = \frac{p}{x \lambda_{L-1,i}} +q
\end{equation*}
so that
\begin{subequations}
\elabel{ql_finite_eps} 
\begin{eqnarray}
\fl \gen{L}{1}(x;\epsilon) \nonumber &&\\
&&\fl =  \epsilon + \sum_{i=0}^{2^{L-1}-1} x \delta \lambda_{L-1,i} 
                  \braket{0}{i}_{L-1} \braket{i}{0}_{L-1} 
  \left(p^2 + q^2 + p q\left(x \lambda_{L-1,i} + \frac{1}{x
 \lambda_{L-1,i}}\right) \right)  \\
&& \fl = \epsilon + \delta\big((p^2 + q^2) x \gen{L-1}{1}(x;\epsilon) 
 + pq (x^2  \gen{L-1}{2}(x;\epsilon) + \gen{L-1}{0}(x;\epsilon) )\big) 
\end{eqnarray}
\end{subequations}
where \Eref{qn_from_evecs} has been used in the last line. Of course,
the generating function $\gen{L}{n}(x;\epsilon)$ is \emph{defined},
\eref{def_q}, for all $\epsilon$ and therefore one can take the limit
$\epsilon\to 0$. This limit should not cause any problems, as $\epsilon$
has only been used to construct the eigenvectors. In fact, the limit
must be identical to setting $\epsilon =0$ in \eref{ql_finite_eps},
as can be shown from \eref{ql_finite_eps} by induction in
$L$. This
finally gives
\begin{equation}\fl
 \gen{L}{1}(x;0) = (p^2 + q^2) x \gen{L-1}{1}(x;0) + pq \left(x^2  \gen{L-1}{2}(x;0) + \gen{L-1}{0}(x;0) \right) \ ,
\end{equation}
where $\gen{L-1}{0}(x;\epsilon)=1$ by \Eref{def_q}, consistent with \Eref{qn_from_evecs}.
In fact, the calculation above can be generalised:
\begin{eqnarray} 
\elabel{qn_generalised}
 \gen{L}{n}(x;\epsilon) & =&  \sum_{i=0}^{2^{L-1}-1} \sum_{j=1}^n \binom{n}{j}
                  \epsilon^{n-j} (x \delta \lambda_{L-1,i})^j \nonumber \\
                  && \times \braket{0}{i}_{L-1} \braket{0}{e_\lambda(\lambda_{L-1,i})}_1  
                           \braket{i}{0}_{L-1} \braket{e_\lambda(\lambda_{L-1,i})}{0}_1  \\
&&+ \sum_{i=0}^{2^L-1} \epsilon^n 
                  \braket{0}{i(x)}_L \braket{i(x)}{0}_L \nonumber
\end{eqnarray}
Again, all sums can be written in terms of $\gen{L-1}{n}(x;\epsilon)$ plus
$\epsilon^n$:
\begin{eqnarray} \elabel{qn_generalised_full}
\fl \gen{L}{n}(x;\epsilon) & = & \epsilon^n + \sum_{j=1}^n \binom{n}{j} \delta^j \epsilon^{n-j}  \nonumber\\
&& \fl  \times
              \Big((p^2+q^2) x^j \gen{L-1}{j}(x;\epsilon) + pq x^{j+1} \gen{L-1}{j+1}(x;\epsilon) + pq  x^{j-1} \gen{L-1}{j-1}(x;\epsilon) \Big) 
\end{eqnarray}
For vanishing dissipation this simplifies to the central result
\begin{equation} \elabel{diffusion_eqn}\fl
 \gen{L+1}{n}(x;0) = x^n \left( \Dtilde \gen{L}{n}(x;0) + D (x \gen{L}{n+1}(x;0) + x^{-1} \gen{L}{n-1}(x;0) )
\right)
\end{equation}
with $D=pq$ and $\Dtilde = p^2+q^2 = 1-2D$. \Eref{diffusion_eqn} is
closely related to a diffusion equation. The boundary conditions are
$\gen{L}{0}(x;\epsilon)\equiv 1$ for $L\ge 1$ as mentioned above and
$\gen{L=0}{n}(x;\epsilon)\equiv 1$. The latter comes from a direct
evaluation of \eref{qn_from_evecs} for $L=1$, which is identical to
\eref{qn_generalised_full} for $\gen{L=0}{n}(x;\epsilon)\equiv 1$.

\subsection{Solving $\gen{L}{n}$}
There is no general solution for \eref{diffusion_eqn} known to the author. However, one can
solve it order by order in derivatives by $x$ at $x=1$, i.e. calculate
every individual moment, see \eref{sm}. In the following, the notation
\begin{subequations}
\begin{eqnarray}
 \gen{L+1}{n} & = & \gen{L+1}{n}(1;0) \\
 \gen{L+1}{n}'& = & \left. \frac{d}{dx}\right|_{x=1} \gen{L+1}{n}(x;0) \ , 
\end{eqnarray}
\end{subequations}
etc. is used. One finds for $n\ge 1$
\begin{equation}
 \gen{L+1}{n} = \Dtilde \gen{L}{n} + D (\gen{L}{n+1} + \gen{L}{n-1})
\end{equation}
which is solved with the boundary conditions introduced above by
$\gen{L}{n}=1$. Of course, this is just normalisation. Using this result
the next derivative is
\begin{equation} \elabel{aves}
 \gen{L+1}{n}' = n + \Dtilde \gen{L}{n}' + D (\gen{L}{n+1}' + \gen{L}{n-1}')
\end{equation}
with the boundary conditions $\gen{L+1}{n=0}'=0$ and
$\gen{L=0}{n}'=0$. The solution of \eref{aves} can easily be guessed as
\begin{equation} \elabel{guess_n_1}
 \gen{L}{n}'=n L \ .
\end{equation}
This is not surprising, because it says that the average number of 
topplings occurring in the system per $n$ kicks is $nL$. That is obviously
true, because every unit added must leave the system by travelling
through the entire lattice.

The next order is the first non-trivial one. The difference equation then
reads 
\begin{equation} \elabel{vars}
 \gen{L+1}{n}'' = (n^2+2D)(2L+1) -n + \Dtilde \gen{L}{n}'' + D (\gen{L}{n+1}'' + \gen{L}{n-1}'') \ .
\end{equation}
Introducing
\begin{equation} \elabel{trans_gen}
\gen{L+1}{n}'' =\src{L+1}{n}+\tgen{L+1}{n}''
\end{equation}
with 
\begin{equation} 
 \src{L}{n}=\sum_{i=0}^{L-1} n^2 - n + 2 i n^2 = -nL + n^2 L^2
\end{equation}
which has the useful property $\src{L+1}{n}-\src{L}{n} = -n + 2 n^2 L +
n^2$ and $\src{L}{n+1}+\src{L}{n-1}=2 \src{L}{n} + 2 L^2$, one finds
after some algebra
\begin{equation}
 \tgen{L}{1}'' = \sum_{l=0}^{L-1} 2D(L-l)^2 \sum_{m=0}^{l} \binom{l}{m} p^{2l-2m} q^{2m} \left( \binom{l}{m} + \binom{l}{m+1} \frac{q}{p}\right) \ .
\end{equation}
In order to analyse the asymptotic behaviour for $L\to\infty$, one
writes
\begin{equation} 
 \tgen{L}{1}'' =  \sum_{l=0}^{L-1} 2D(L-l)^2 \phi^*(l)
\end{equation}
with
\begin{equation}
 \phi^*(l) = \sum_{m=0}^l \binom{l}{m} p^{2(l-m)} q^{2m} \left( \binom{l}{m} + \binom{l}{m+1} \frac{q}{p}\right)\ .
\end{equation}
The binomials, which can be approximated by a Gaussian, are treated
identically and the summation can be written as an integral,
so that finally 
\begin{equation}
 \tgen{L}{1}'' \to \int_0^{L-1} 2D (L-l)^2  \frac{1}{\sqrt{p q \pi}} \left( \erf(\sqrt{lp/q}) + \erf(\sqrt{lq/p})\right)
\end{equation}
where $\erf(x)\equiv 2 \int_0^x dz \exp(-z^2)/\sqrt{\pi}$. In leading
order, this turns out to be 
\begin{equation} \elabel{ave2_leading}
 \tgen{L}{1}'' \to \frac{32}{15 \sqrt{\pi}} \sqrt{p q }L^{5/2} \ ,
\end{equation}
which is according to \eref{trans_gen} also the leading order of
$\gen{L+1}{n}''$ and therefore the leading order of $\ave{s^2}$, see
\eref{sm} with \eref{guess_n_1}.
This is perfectly confirmed by numerical simulations of the model.

The two exponents $\gamma_1=1$ (see \Eref{guess_n_1}) and
$\gamma_2=5/2$ \eref{ave2_leading} lead together with \eref{gamma_n}
to $\tau=4/3$ and $D=3/2$.

\section{Reaction-Diffusion mapping} \label{sec:reaction_diffusion}
It is possible to map the model onto a very simple reaction-diffusion
process of the form $A+A\to A$ \cite{benAvrahamHavlin:2000}. To this
end, the configuration of the lattice is described by the thick line
shown in \Fref{react_diff}. The line consists of segments, which can
either point up or down by an angle of $45^\circ$. If the line corresponding to the $i$th
site goes up, it indicates that the $i$th site is in state $z=1$,
otherwise the line goes down indicating the state of the site to be
$z=2$. According to \eref{stat_dist} the configuration of the lattice
(in the stationary state)
after an avalanche is a product state, where a site is in state $z=1$
with probability $p$ and in state $z=2$ with probability $q$. Thus, the
thick line is in fact the trajectory of a random walker with drift
corresponding to the difference $p-q$.

\begin{figure}[th]
\begin{center}
\newpsobject{showgrid}{psgrid}{subgriddiv=1, griddots=10,gridlabels=6pt}
\begin{pspicture}(-1,-0.5)(6,6) 

\newcommand{\nsize}{0.16}

\pscircle(-1,1){\nsize}
\pscircle(-1,3){\nsize}

\pscircle(0,0){\nsize}
\pscircle(0,2){\nsize}
\pscircle(0,4){\nsize}

\pscircle(1,1){\nsize}
\pscircle(1,3){\nsize}
\pscircle(1,5){\nsize}

\pscircle(2,0){\nsize}
\pscircle(2,2){\nsize}
\pscircle(2,4){\nsize}

\pscircle(3,1){\nsize}
\pscircle(3,3){\nsize}
\pscircle(3,5){\nsize}

\pscircle(4,0){\nsize}
\pscircle(4,2){\nsize}
\pscircle(4,4){\nsize}

\pscircle(5,1){\nsize}
\pscircle(5,3){\nsize}
\pscircle(5,5){\nsize}

\pscircle(6,0){\nsize}
\pscircle(6,2){\nsize}
\pscircle(6,4){\nsize}

\psline[linewidth=3pt](-1,1)(0,2)(1,1)(2,2)(3,3)(4,2)(5,3)(6,2)

\psline[linewidth=1.5pt,linestyle=dashed](-1,3)(0,4)(1,5)(2,4)(3,3)

\psline(-1,1)(-1.7,1)
\psline(-1,3)(-1.7,3)

\psline[linestyle=dotted](-0.5,-0.4)(-0.5,5.5)
\psline[linestyle=dotted]( 0.5,-0.4)( 0.5,5.5)
\psline[linestyle=dotted]( 1.5,-0.4)( 1.5,5.5)
\psline[linestyle=dotted]( 2.5,-0.4)( 2.5,5.5)
\psline[linestyle=dotted]( 3.5,-0.4)( 3.5,5.5)
\psline[linestyle=dotted]( 4.5,-0.4)( 4.5,5.5)
\psline[linestyle=dotted]( 5.5,-0.4)( 5.5,5.5)
\put(-0.6,-0.75){$1$}
\put( 0.4,-0.75){$2$}
\put( 1.4,-0.75){$3$}
\put( 2.4,-0.75){$4$}
\put( 3.4,-0.75){$5$}
\put( 4.4,-0.75){$6$}
\put( 5.4,-0.75){$7$}

\put(-0.75,1.55){$p$}
\put( 0.6,1.55){$q$}

\psline{<->}(-1.5,1)(-1.5,3)
\put(-1.9,2){$a_1$}

\end{pspicture}
\end{center}
\caption{\flabel{react_diff} The thick, full line shows the
 configuration of the lattice after an avalanche has passed
 through. Each up or down-pointing segment corresponds to a single site,
 the position label of which is shown under the dotted line. A segment pointing
 upwards corresponds to a site being in state $z=1$ (with probability
 $p$, see \Eref{stat_dist}), a segment pointing downwards corresponds to
 state $z=2$ (probability $q$), as indicated. The dashed line
 corresponds to a ``toppling trajectory'' as explained in the text.
}
\end{figure}
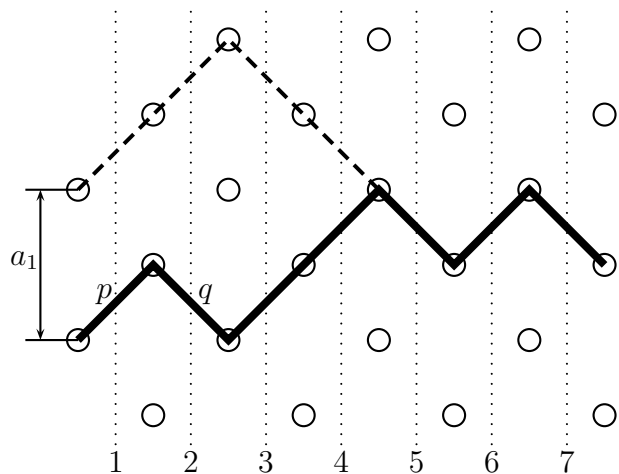

The avalanche itself, on the other hand, is a random walk with the same probabilities. One
can see that by considering the activity $a_i$, which is the number
charges received at site $i$ during an update-sweep as described in
Sec.~\ref{sec:abelian}. From site to site, the activity can either remain constant or
change by $1$ up or down. Apparently, $a_1=1$ is the driving. If a site
receives $a_i$ charges and changes state by $\Delta z_i = z_i(t)-z_i(t+1)$,
then its right neighbour receives $a_{i+1}=a_i+\Delta z_i$ charges. If $\Delta
z_i=0$, then the vertical distance between two consecutive configuration
trajectories (as shown as thick and dashed lines in \Fref{react_diff})
does not change. If, however, the new configuration of site $i$
has an increased value $z_i(t+1)>z_i(t)$, the activity goes down,
because $\Delta z_i<0$. The
only way to increase $z_i$ is to go from state $1$ to state $2$,
i.e. the line segment of the former configuration points up (probability
$p$), while the line segment of the new configuration points down
(probability $q$), so that the gap between the two trajectory
decreases, see \Fref{react_diff} at the dotted line $3$. Similarly, if
the activity goes up, then $\Delta z_i>0$ and the gap increases.

After the activity vanishes, the profile of the new configuration
remains unchanged compared to the former, i.e. the gap between the two
configurations is a constant. In fact, if the gap was initially $1$ and
goes up and down by $1$ as described above, then the gap will be $0$ as
soon as the activity vanishes. This is exactly what is shown in
\Fref{react_diff}: \emph{The thick dashed line shows the new
configuration and its distance to the old configuration is the activity
during the avalanche.} This avalanche occurs within the configuration
shown as a thick line, initiated by a single kick. Initially the gap is
$a_1=1$. If the dashed line would go down immediately on site $1$, the
site would have ``absorbed'' the initial unit and would be in state
$z=2$ (i.e. a segment pointing down). Instead, in the example, it goes
up twice; first just like in the old configuration so that the activity
does not increase, and then in the opposite direction to the old
configuration so that the activity increases by $1$. On site $3$ and $4$
it goes down twice; the toppling on site $4$ is particularly
interesting. Here, initially the activity is $1$, i.e. the site has
received one unit. But the site is in state $z=1$, so it absorbs the
unit with probability $q$, corresponding to the probability of the
dashed line segment to point downwards.

The activity is measured half a unit left of each site as the distance
between old and new trajectories, which, in turn, is measured in such
units that the vertical distance between two circles (in
\Fref{react_diff}) is $1$. The
reason for the shift is that one wants to measure how many charges have 
\emph{arrived} at a site, not affected by the value of the resulting activity.

To repeat this important point, the trajectory of an avalanche becomes
the configuration for the next avalanche, i.e. the thick dashed line in
\Fref{react_diff} becomes the thick solid line for the next avalanche.

One can calculate the probability of the changes of activity explicitly:
The new segment goes up with probability $p$ and down with probability
$q$, the same applies to the old segment. Thus, they point in the same
direction (no change of activity) with probability $p^2+q^2$, the gap
widens with probability $p q$ and shrinks with $q p$. Hence, the gap
between the two trajectories is in fact a symmetric random walk, even
though the individual trajectories might have a bias, according to
$p-q$. 

As described above (see Sec.~\ref{sec:model}), the avalanche size is
measured as the number of topplings. For convenience, one can define it
as the number of charges, which makes hardly any difference, because the
number of topplings of site $i$ is identical to the number of charges on
site $i+1$, unless $i=L$, simply because there is no site $L+1$;
similarly for $L=1$.  

In the following, we will consider the
number of charges as the avalanche size, because the total number of
charges is simply the area between two of those trajectories described
above, namely the sum over all activities. From this it is also clear that
the avalanche size \emph{is} actually uniquely determined by the initial
and the final configuration, with initial activity $a_1=1$.

\subsection{Relation to other models}
Before the above identification of the process as a random walker is
cast into an continuum problem and subsequently solved, it is worth
pointing out other models which are closely linked to the present
one. 

\subsubsection{Anisotropic BTW model}
Dhar and Ramaswamy \cite{DharRamaswamy:1989} developed an anisotropic
variant of the well-known BTW sandpile model \cite{BakTangWiesenfeld:1987},
which is now known as the directed sandpile model. This model, however, is
situated on a $1+1$-dimensional lattice and the annihilating random
walkers represent the contours of the compact area covered by an
avalanche. The randomness here comes solely from the randomness of whether
a site charged by particles from toppling sites topples in turn. An
equivalence to a variant of directed percolation has already been
pointed out in \cite{DomanyKinzel:1984}, see also \cite{MohantyDhar:2002}.

Kloster, Maslov and Tang \cite{KlosterMaslovTang:2001} have studied a
stochastic directed sandpile model, which was originally proposed by
Pastor-Satorras and Vespignani
\cite{Pastor-SatorrasVespignani:2000a}. This model is closely related to
the one presented in this paper, even though it is also situated
on a $1+1$-dimensional lattice. The
authors find the same exponents by scaling arguments. The mapping to the
two-dimensional reaction-diffusion process presented above, questions
their assertion that their model is in a different universality class
than the model by Dhar and Ramaswamy.

For these models it is fairly obvious how to extend them systematically
to higher dimensions. Using scaling arguments in conjunction with some
simplifying assumptions, Paczuski and Bassler
\cite{PaczuskiBassler:2000} arrive at a general expression for the value
of the exponents of this model in higher dimensions. Unfortunately, it
is not so clear how to generalise the model studied in this paper to higher
dimensions, because it is unclear how to generalise the driving and what
boundary conditions to apply.

\section{Continuum solution}
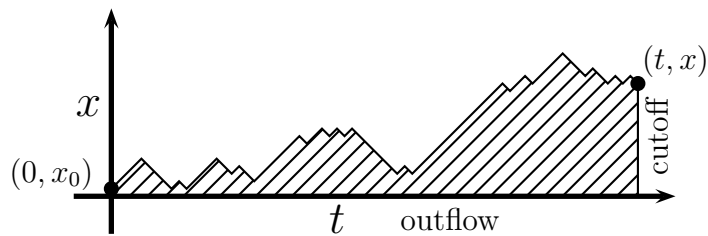
\begin{figure}[th]
\begin{center}
\newpsobject{showgrid}{psgrid}{subgriddiv=1, griddots=10,gridlabels=6pt}
\begin{pspicture}(-1.0,-1.0)(7,2.5) 
\psset{xunit=0.1cm,yunit=0.1cm}
\input{area_under_rw}
\end{pspicture}
\caption{\flabel{area_under_rw} The area under the trajectory (hatched) is the
 avalanche size. The two filled circles mark the starting point
 $(0,x_0)$ and the end point $(t,x)$}
\end{center}
\end{figure}
Having mentioned already the mapping to an annihilating random walk, the
continuum description is straight forward. To this end, the quantity
$\psi_n(t,x;x_0)$ is introduced. It quantifies the properties of a random
walker along an absorbing wall. For $n=0$ it is the probability density
of random walkers at time $t$ and height $x$ over the absorbing wall,
starting at height $x_0$, which is $x_0=1$ for a single initial kick. Here, $t$ takes
on the r{\^o}le of the horizontal (continuous) position between $t=0$
and $t=L$
in a picture like \Fref{react_diff}. To motivate the following
calculation, one imagines a large set of trajectories of random walkers along
the absorbing wall from $t=0$, $x=x_0$ to $t$ and $x$. The set of areas under the trajectories,
as exemplified in \Fref{area_under_rw}, is then
$\{s_i(t,x;x_0)\}$, where $i$ is indexing the elements in the set. $\aves{s^n(t,x;x_0)}$
is the average of the $n$th moment over this set. Now one can express
the time-evolution of this average as the sum of three contributions of
the three processes of up, down or straight movement of the random
walker. Thus, up to terms of order $\Delta t \Delta x$ (see caption of
\Fref{each_process})
\begin{subequations}
\elabel{m_eq}
\begin{eqnarray}
     \psi_0(t+\Delta t,x;x_0)  \aves{s^n(t+\Delta t,x;x_0)}  && \nonumber \\
= pq          \psi_0(t,x+\Delta x;x_0) \aves{ (s(t,x+\Delta x;x_0) + x \Delta t)^n} && \\
\quad + (p^2 + q^2) \psi_0(t,x;x_0)          \aves{ (s(t,x;x_0) + x \Delta t)^n}          && \\
\quad + pq          \psi_0(t,x-\Delta x;x_0) \aves{ (s(t,x-\Delta x;x_0) + x \Delta t)^n} && \elabel{from_below}
\end{eqnarray}
\end{subequations}
where each term corresponds to a process like the one shown in
\Fref{each_process}. The multiplication by $\psi_0(t,x;x_0)$ is
necessary in order to weight each of the ensembles for each contribution
properly. For example, there might a much larger contribution from below,
even though on average the moment at this position is smaller than at
the other positions.

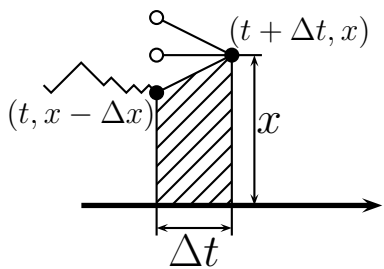
\begin{figure}[th]
\begin{center} 
\newpsobject{showgrid}{psgrid}{subgriddiv=1, griddots=10,gridlabels=6pt}
\begin{pspicture}(0.0,-1.0)(3,3) 
\input{each_process}
\end{pspicture}
\caption{\flabel{each_process} A new segment (hatched area) is added to
 the currently considered path, increasing all areas in the ensemble
 $\{s_i(t,x -\Delta x;x_0)\}$ by $x \Delta t + \OC(\Delta t \Delta x)$
 and producing a new ensemble $\{s_i(t+\Delta t,x;x_0)\}$. The example
 shown corresponds to \eref{from_below}, which starts at $(t,x-\Delta
 x)$. Starting points of other contributions are shown as empty
 circles. The coordinates of the two black points are given in the form
 $(t,x)$.}
\end{center}
\end{figure}

Defining 
\begin{equation} \elabel{def_psi}
 \psi_n(t,x;x_0) \equiv  \psi_0(t,x;x_0) \aves{ s(t,x;x_0)^n }
\end{equation}
one finds in the continuum limit of \eref{m_eq} (keeping $D \Delta t/\Delta
x^2$ constant)
\begin{equation} \elabel{pde}
  \partial_t \psi_n(t,x;x_0) = D \partial_x^2 \psi_n(t,x;x_0) + x n \psi_{n-1}(t,x;x_0)
\end{equation}
where $D=p q$ again\footnote{It is interesting to note that this can be
written using a generating function $\Psi(t,x;x_0,\xi)$ with 
$\partial_t \Psi=x \xi \Psi + D \partial_x^2 \Psi$, so that indeed
$\left. \frac{d^n}{d \xi^n} \right|_{\xi=0} \Psi(t,x;x_0,\xi) = \psi_n(t,x;x_0)$
}. The boundary conditions for $n=0$ are observed immediately and
transfered to $\psi_n$ using \eref{def_psi} by noting that
$\aves{s(t,x;x_0)^n}$ is non-divergent, so 
\begin{subequations}
\begin{eqnarray}
 \lim_{t\to 0} \psi_n(t,x;x_0) & = & \delta_{n,0} \delta(x-x_0) \\
 \psi_n(t,0;x_0) & = & 0 \elabel{absorbing}
\end{eqnarray}
\end{subequations}
and the PDE \eref{pde} is to be solved for $x \in [0, \infty[$. 

The avalanche sizes are measured from avalanche trajectories which have died
out or reached the end of the system. Thus, the averages measured in the
model are taken from the random walkers which have reached the
absorbing wall or did not do so until a cutoff time $t$. Therefore the $n$th
moment observed is
\begin{equation}
 \ave{s^n}(t;x_0) = \int_0^t dt' j_n(t';x_0) + \int_0^\infty dx' \psi_n(t,x';x_0)
\end{equation}
where the first integral runs over the ``outflow'', $j_n(t,x=0;x_0)
\equiv -D \partial_x|_{x=0} \psi_n(t,x;x_0)$ and the second over the
contributions at cutoff time (see marks in
\Fref{area_under_rw}). $\ave{s^n}(t;x_0)$ denotes the $n$th moment of
the avalanche size (measured as the number of charges) for a system of
size $t$ starting with $x_0$ initial charges.  Using \eref{pde} one has
\begin{equation} \elabel{sn_from_psi}
 \ave{s^n}(t;x_0) = \int_0^t dt' \int_0^\infty dx' x' n \psi_{n-1}(t',x';x_0)
\end{equation}
The dimensionless form of $\psi$ is given by 
\begin{equation} \elabel{def_wp}
\psi_n(x,t;x_0) = \frac{1}{x_0} \left(\frac{x_0^3}{D}\right)^n \WP_n(y,\tau) 
\end{equation}
with
$y=x/x_0$ and $\tau=t/(x_0^2/D)$. The propagator $G(y,\tau;y_0)$ is
easily obtained from a mirror-charge trick,
\begin{equation}
 G(y,\tau;y_0) \equiv \frac{1}{\sqrt{4 \tau \pi}} \left( 
e^{\frac{y-y_0}{4\tau}} - e^{\frac{y+y_0}{4\tau}}
\right) \ ,
\end{equation}
and $\WP_0(y,\tau)=G(y,\tau;1)$, i.e.
\begin{equation}
 \WP_0(y,\tau)=\frac{1}{\sqrt{\tau \pi}} e^{-\frac{y^2+1}{4 \tau}} \sinh\left(\frac{y}{2\tau}\right)
\ .
\end{equation}
One might be inclined to transfer the problem into $k$-space, which, however, does not
simplify the problem because of the boundary condition
\eref{absorbing}. The expression
\begin{equation}
 \WP_n(y,\tau) = \int_0^\tau d\tau' \int_0^\infty dy' n y' \WP_{n-1}(y',\tau') G(y,\tau-\tau';y') 
\end{equation}
is the formal solution. Rescaling the arguments of $\WP_n$ by powers of $\mu$ one finds
\begin{equation}
 \WP_n(\sqrt{\mu} y,\mu \tau) = 
\mu^{3/2} \int_0^\tau d\tau' \int_0^\infty dy' n y' \WP_{n-1}(\sqrt{\mu} y',\mu \tau') G(y,\tau-\tau';y') 
\ .
\end{equation}
From $\WP_{n-1}(\sqrt{\mu}y, \mu \tau)=\mu^{\alpha_{n-1}} \WP_{n-1}(y,\tau)$,
then follows $\WP_n(\sqrt{\mu}y, \mu \tau)=\mu^{\alpha_{n-1}+3/2}
\WP_n(y,\tau)$. Thus, starting with $\WP_0(\sqrt{\mu}y, \mu
\tau)=\mu^{\alpha_0} \WP_0(y,\tau)$ one has
apparently
\begin{equation} \elabel{scaling}
 \WP_n(\sqrt{\mu} y,\mu \tau) = \mu^{\frac{3}{2}n + \alpha_0}  \WP_n(y,\tau) \ .
\end{equation}
Unfortunately the scaling behaviour of $\WP_0$ is a bit more
complicated. Nevertheless, it can be expanded for large $\mu$, or
actually large $\mu \tau$, as
\begin{equation} \elabel{expansion_WP0}
 \WP_0(\sqrt{\mu} y,\mu \tau) = \frac{1}{\mu} \frac{1}{\sqrt{\tau \pi}} e^{-\frac{y^2}{4\tau}}
\left(\frac{y}{2\tau} + \frac{1}{\mu} \left(\frac{y^3}{48 \tau^3}-\frac{y}{8 \tau^2} \right)
+ \dots \right) \ .
\end{equation}
Bearing in mind the necessity of large $\mu \tau$ one can now apply the
scaling argument \eref{scaling} order by order in $\mu$ since \Eref{pde}
and its dimensionless counterpart are
linear. From \eref{expansion_WP0} it is $\alpha_0=-1$ for the leading
order, $\alpha_0=-2$ for the first sub-leading order and so on.

\Eref{scaling} immediately translates to $\ave{s^n}$ using
\eref{sn_from_psi} and \eref{def_wp}; to leading order one finds
\begin{equation} \elabel{aves_scaling}
\ave{s^n}(\mu t;x_0) = \mu^{(3/2) n + 1/2 +
\alpha_0}\ave{s^n}(t;x_0) + \dots
\end{equation}
\emph{Assuming} \eref{def_expos}, from \eref{moment_scaling} with $t$
taking the r{\^o}le of $L$ it follows that $D=3/2$ and
$D(1-\tau)=1/2+\alpha_0$, i.e. for $\alpha_0=-1$ one has $\tau=4/3$. The
next order correction is $D=3/2$ and $\tau'=2$.

Of course, it is also possible to calculate the leading orders of
$\ave{s^n}$ exactly. Because of \eref{aves_scaling}, one needs to
calculate $\ave{s^n}(\mu t;x_0)$ for one value of $t$ only. The 
simplest choice is to set $t=x_0^2/D$, which gives 
$\ave{s}(x_0^2/D;x_0) = x_0^3/D$ for $n=1$, i.e.
\begin{equation} \elabel{ave_s_cont}
 \ave{s}(t;x_0) = x_0 t
\end{equation}
which is exactly \eref{guess_n_1} ($n$ in \eref{guess_n_1} corresponds
to $x_0$ here and $L$ in \eref{guess_n_1} to $t$). This is actually
surprising, because \eref{ave_s_cont} is only the \emph{leading} order and
corrections are expected from higher orders. However, it turns out
that in fact all higher order corrections cancel. Indeed, remarkably
\begin{equation}
 \int_0^\infty dx x e^{-\frac{x^2}{4t}} \left(\frac{x}{t} - 
	2 e^{-\frac{1}{4t}} \sinh\left(\frac{x}{2 t}\right)\right) = 0 \ .
\end{equation}
even though $x/t$ is only the leading order of $2
\exp(-1/(4t))\sinh(x/(2 t))$. Especially
\begin{equation}
 \int_0^\infty dx x e^{-\frac{x^2}{4t}} 
\left(\frac{x^3}{48 t^3}-\frac{x}{8 t^2} \right) = 0 \ .
\end{equation}

According to \eref{sn_from_psi} the next moment is
\begin{equation}
 \ave{s^2}(\mu t;x_0) =
2 \mu^{5/2} \int_0^{tD/x_0^2} d\tau \int_0^\infty dy \frac{x_0^3}{D} y \WP_1(y,\tau)
\end{equation}
the leading order of which can be determined using the leading order of
$\WP_1$, 
\begin{equation}\fl
  \WP_1(y,\tau) =
\int_0^\tau d\tau' \int_0^\infty dy'
\frac{1}{\sqrt{\tau' \pi}}
e^{-\frac{y'^2}{4 \tau'}} \frac{y'^2}{2 \tau'}
\frac{1}{\sqrt{4 \pi (\tau -\tau')}}
\left(
e^{-\frac{y-y'}{4(\tau-\tau')}} - e^{-\frac{y+y'}{4(\tau-\tau')}}
\right) + \dots
\end{equation}
which gives the leading order of $\ave{s^2}$
\begin{equation}
 \ave{s^2}(t;x_0) = \frac{32}{15 \sqrt{\pi}} t^{5/2} \sqrt{D x_0^2} + \OC(t^{3/2})
\end{equation}
identical to \eref{ave2_leading}. Higher orders become very
tedious, so that numerical evaluation seems to offer the better option.

\section{Discussion and Conclusion}
The results above represent some of the few exact result for
sandpile-like models:
\Eref{guess_n_1} and \Eref{ave2_leading} are the exact leading orders of
the first two moments of the avalanche size distribution without making
any assumptions about scaling behaviour. The conclusion that $\tau=4/3$
and $D=3/2$ can only be drawn by either assuming \eref{def_expos}, or by
accepting the continuum result \eref{aves_scaling} and 
using 
the uniqueness of the distribution inferred from its
moments.\footnote{Most crucially, the moment generating function $\sum
x^n \ave{s^n}/n!$ must actually exist, which is at least reasonable to
assume from \eref{aves_scaling}.}

The method introduced in Sec.~\ref{sec:markov} is not restricted to
sandpile-like models. The underlying idea is to use a Markov matrix not
only to evolve the state distribution, but also to calculate the moment
generating function of the relevant observable. In order to obtain the
finite-size scaling behaviour, its set of eigenvectors is generated
recursively. From this recursion relation one can then develop a
(discrete) PDE like \eref{diffusion_eqn}, which can subsequently be used
as a starting point for other techniques. In a two-dimensional variant
of the present model, this recursion relation is much more complicated
to obtain and might require the use of a matrix product state ansatz
\cite{DerridaEvans:1997}.  Nevertheless, it seems promising to apply the
approach to more complicated processes, such as the TASEP
and recent
variants \cite{ParmeggianiFranoschFrey:2003}, for which there is no
solution known yet.

\ack{The author gratefully acknowledges the support of EPSRC and the
hospitality of Holger Bruhn, Inge Ruf and Valentine Walsh. The author
wishes to thank Simone Avogadro di Collobiano and Henrik Jensen for very
helpful discussions and Nicholas R. Moloney for proofreading.}

\section*{References}
\bibliography{articles,books}
\end{document}

%% file: area_under_rw.tex
\pspolygon[fillstyle=vlines,hatchangle=-45,fillcolor=black](0,0)%
(0,1)%
(1,2)%
(2,3)%
(3,4)%
(4,5)%
(5,4)%
(6,3)%
(7,2)%
(8,1)%
(9,2)%
(10,1)%
(11,2)%
(12,3)%
(13,4)%
(14,5)%
(15,4)%
(16,3)%
(17,4)%
(18,3)%
(19,2)%
(20,3)%
(21,4)%
(22,5)%
(23,6)%
(24,7)%
(25,8)%
(26,7)%
(27,8)%
(28,9)%
(29,8)%
(30,9)%
(31,8)%
(32,9)%
(33,8)%
(34,7)%
(35,6)%
(36,5)%
(37,4)%
(38,3)%
(39,4)%
(40,3)%
(41,4)%
(42,5)%
(43,6)%
(44,7)%
(45,8)%
(46,9)%
(47,10)%
(48,11)%
(49,12)%
(50,13)%
(51,14)%
(52,15)%
(53,14)%
(54,15)%
(55,16)%
(56,15)%
(57,16)%
(58,17)%
(59,18)%
(60,19)%
(61,18)%
(62,17)%
(63,16)%
(64,17)%
(65,16)%
(66,15)%
(67,16)%
(68,15)%
(69,16)%
(70,15)%
(70,0)
\psline[linewidth=2pt]{->}(-5,0)(75,0)
\psline[linewidth=2pt]{->}(0,-5)(0,25)
\rput(30,-3){{\Large $t$}}
\rput(-3,12){{\Large $x$}}

\rput(45,-3){outflow}
\rput(72.5,8){\rotatebox{90}{cutoff}}

\pscircle[fillstyle=solid,fillcolor=black](70,15){0.1cm}
\rput(75,18){$(t,x)$}
\pscircle[fillstyle=solid,fillcolor=black](0,1){0.1cm}
\rput(-8,3){$(0,x_0)$}

%% file: each_process.tex
\psline[linewidth=2pt]{->}(0,0)(4,0)
\psline(1,-0.4)(1,0)
\psline(2,-0.4)(2,0)
\psline(2,2)(2.4,2)
\rput(2.5,1.1){{\Large $x$}}
\rput(1.5,-0.6){{\Large $\Delta t$}}
\pspolygon[fillstyle=vlines,hatchangle=-45,fillcolor=black](1,0)(1,1.5)(2,2)(2,0)

\psline(1,2.5)(2,2)
\pscircle[fillstyle=solid](1,2.5){0.1cm}
\psline(1,2)(2,2)
\pscircle[fillstyle=solid](1,2){0.1cm}
\pscircle[fillstyle=solid,fillcolor=black](1,1.5){0.1cm}
\rput(0.0,1.2){$(t,x-\Delta x)$}
\pscircle[fillstyle=solid,fillcolor=black](2,2){0.1cm}
\rput(2.9,2.3){$(t+\Delta t,x)$}

\psline{<->}(1,-0.3)(2,-0.3)
\psline{<->}(2.3,0.0)(2.3,2)
\psset{xunit=0.1cm,yunit=0.1cm}
\rput(-60,0){
\psline(55,16)%
(56,15)%
(57,16)%
(58,17)%
(59,18)%
(60,19)%
(61,18)%
(62,17)%
(63,16)%
(64,17)%
(65,16)%
(66,15)%
(67,16)%
(68,15)%
(69,16)%
(70,15)
}